\documentclass[conference]{IEEEtran}
\IEEEoverridecommandlockouts
\usepackage{cite}
\usepackage{amsmath,amssymb,amsfonts}
\usepackage{algorithmic}
\usepackage{graphicx}
\usepackage{textcomp}
\usepackage{xcolor}
\usepackage{url}
\def\BibTeX{{\rm B\kern-.05em{\sc i\kern-.025em b}\kern-.08em
    T\kern-.1667em\lower.7ex\hbox{E}\kern-.125emX}}

\usepackage{fancyhdr}
\newcommand{\revision}[1]{{\color{black}#1}}
\begin{document}

\title{Drawing the Line: Where AI Guidance and Human Creativity Meet in Emotion-Driven Comic Storyboarding\\


\author{
\IEEEauthorblockN{1\textsuperscript{st} Jocelyn Shen\textsuperscript{*}}
\IEEEauthorblockA{\textit{MIT Media Lab}\\
Cambridge, MA, USA\\
joceshen@mit.edu}
\and
\IEEEauthorblockN{2\textsuperscript{nd} Isabella Pu\textsuperscript{*}}
\IEEEauthorblockA{\textit{MIT Media Lab}\\
Cambridge, MA, USA\\
ipu@mit.edu}
\and
\IEEEauthorblockN{3\textsuperscript{rd} Alessandro Brise\~{n}o}
\IEEEauthorblockA{\textit{MIT}\\
Cambridge, MA, USA\\
jediale@mit.edu}
\and
\IEEEauthorblockN{4\textsuperscript{th} Hyun Kim}
\IEEEauthorblockA{\textit{MIT}\\
Cambridge, MA, USA\\
hkim2022@mit.edu}
\and
\IEEEauthorblockN{5\textsuperscript{th} Fiona Lu}
\IEEEauthorblockA{\textit{MIT}\\
Cambridge, MA, USA\\
froit@mit.edu}
\and
\IEEEauthorblockN{6\textsuperscript{th} Sharifa Alghowinem}
\IEEEauthorblockA{\textit{MIT Media Lab}\\
Cambridge, MA, USA\\
sharifah@mit.edu}
\and
\IEEEauthorblockN{7\textsuperscript{th} Cynthia Breazeal}
\IEEEauthorblockA{\textit{MIT Media Lab}\\
Cambridge, MA, USA\\
cynthiab@mit.edu}
\and
\IEEEauthorblockN{8\textsuperscript{th} Hae Won Park}
\IEEEauthorblockA{\textit{MIT Media Lab}\\
Cambridge, MA, USA\\
haewon@mit.edu}
}
\thanks{*Authors contributed equally.}
}

\maketitle
\thispagestyle{fancy}

\begin{abstract}

While generative AI models can produce visually faithful artwork, they often fall short in conveying emotional authenticity---a key driver of human expression. In visual storytelling, particularly comic storyboarding, this gap becomes pronounced: effective storyboards require both \textit{technical} knowledge (e.g., anatomical accuracy or scenic composition) and \textit{emotional insight} (from lived experience). We explore how human-AI collaboration can support emotion-driven creativity, where the user's feelings guide generation and emotional resonance is the goal. We present \textsc{EmoToon}, a technology probe that helps non-professional artists generate sketch-like storyboards and iterate on visual ideas. In a controlled study with $N=25$ participants, we find that AI assistance significantly improves emotional expression, aesthetic quality, and exploration, but reduces users' creative ownership. 
Our findings offer broader insights for human-AI co-creation in the domain of comic storytelling, emphasizing balance between output quality and user freedom, and raising new challenges for image generation models in this domain.

\end{abstract}


\maketitle

\begin{figure*}[htbp]
\centering
  \includegraphics[width=.7\textwidth]{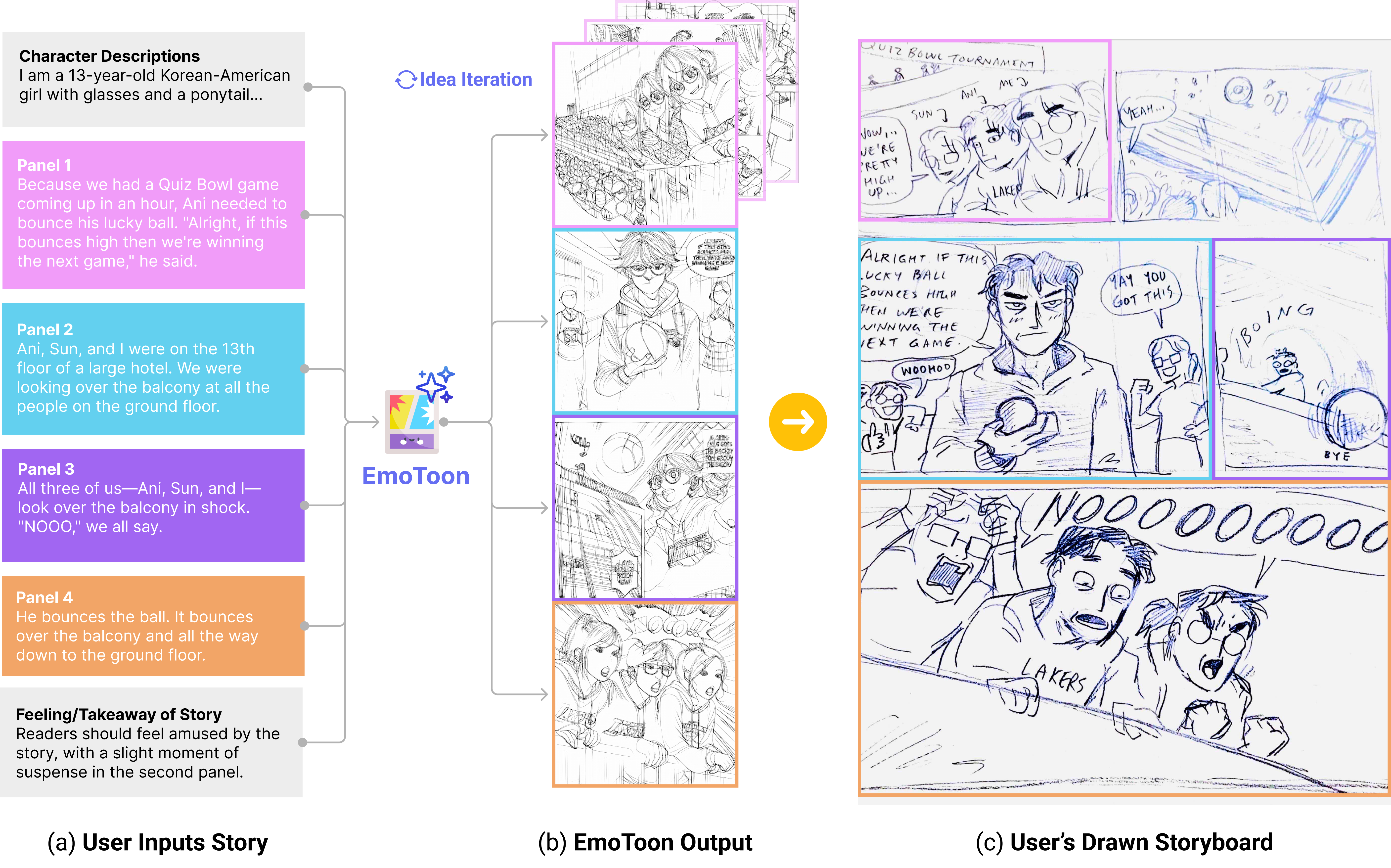}
  \caption{We implement a technology probe to better understand the dichotomy between using generative AI to guide artistic \textit{skill} versus allowing room for human \textit{authenticity} in conveying emotions.
  EmoToon is an AI-driven prototype that empowers non-professional storytellers to create comic storyboards based on the emotions they hope to convey through their story. Textual inputs are converted into drawing inspirations, where ideas can be regenerated and revisited for iteration during the user's drawing process. }
  \label{fig:teaser}
  \vspace{-15pt}
\end{figure*}

\section{Introduction}
 With current advances in generative artificial intelligence (AI), models are capable of outputting highly realistic images and videos in a variety of artistic styles. However, many artists question the authenticity of AI-generated art in being able to convey the human experience \cite{ali_studying_2023}. In particular, empirical studies find that AI-generated art evokes \textit{less emotion and narrative} than human-made art \cite{bellaiche_humans_2023}. Foundational creativity research repeatedly observes that centering emotions yields more creative outcomes, highlighting that emotion is both a \textit{generative} and \textit{evaluative} component of creativity---artists draw on their own emotions to create art, and viewers as well as artists themselves appreciate art for the ways in which a piece emotionally resonates \cite{de_dreu_hedonic_2008, gonzalez-zamar_emotional_2021}. While models may excel in {knowledge-intensive tasks}, such as anatomical accuracy when generating a character \cite{xu_expertprompting_2023, petroni_language_2019}, they cannot replace or replicate the rich array of emotions in people's lived experiences.


 A central challenge for AI in the creative domain is to understand the tension between \textit{technical sophistication} of generative models and \textit{emotional authenticity} of human storytelling, such that human-AI co-creation best leverages the ``superpowers'' of both people and AI systems.
In the human-computer interaction (HCI) community, many researchers have identified human-AI collaboration as an opportunity to help people co-create art in the storyboarding process \cite{zhang_storydrawer_2022, shi_emog_2020}, but few have explicitly explored how AI systems can support \textbf{emotion-driven storytelling}. We define emotion-driven creativity as a process where the user's feelings guide the creative generation (input), and emotional resonance becomes a key goal of the output \cite{gonzalez-zamar_emotional_2021}.
\textbf{In this work, we explore how generative AI tools can be used to augment emotional expression in comic storyboarding, using a technology probe and empirical study to understand the dichotomy between providing enough technical guidance to non-professional artists and preserving human emotional authenticity.}

\revision{We scope our work specifically to non-professional artist participants as this population stands to benefit the most from AI-assisted visual storytelling. Non-experts frequently possess rich, emotionally resonant lived experiences and narratives they wish to express, yet they typically lack the formal technical vocabulary, execution skills, or training in visual design necessary to translate those feelings into robust visual assets. Focusing on this user demographic allows us to investigate how human-AI co-creation can make emotionally driven visual storytelling highly accessible to everyday creators.}

To this end, our paper presents the following contributions:
\begin{itemize}
    \item We developed \textsc{EmoToon}, a technology probe that helps non-expert storytellers construct comic storyboards based on the feeling they are trying to express in their narrative.
    \item We conducted a user study with $N=25$ participants to assess users' self-perceived impact of generative AI assistance on (1) conveying the storyteller's emotional content, (2) aesthetic quality of the user's storyboard, (3) level of creative freedom, and (4) ease of learning for non-expert artists  when using \textsc{EmoToon} compared to freehand drawing.
\end{itemize}

\section{Related Work}

\subsection{Human Perceptions on Creative AI}

As AI increasingly permeates creative domains, its use in fields such as painting \cite{xu2023everyone,liu_when_2024,sun2022pigments} and music \cite{louie2020novice,suh2021ai} has attracted growing attention. These tools enable creators to explore new forms of expression that blend traditional artistic practices with computational capabilities, and some observers perceive AI-generated works as approaching human-level artistry \cite{hong2018bias}. At the same time, evolving technologies also reshape cultural understandings of art \cite{sun2022pigments}, and many artists remain hesitant to fully embrace AI systems. Interviews with artists reveal concerns that AI art relies on unconsented training data, lacks authenticity, and misses the human creativity that defines artistic practice \cite{ali_studying_2023}.

The broader public expresses similar skepticism \cite{latikka_ai_2023,oksanen_artificial_2023,horton_jr_bias_2023,bower_perceptions_2021}. People often rate AI-generated artworks lower than human-made ones even when unable to distinguish between them \cite{oksanen_artificial_2023}. When images are labeled as ``AI-made,'' viewers perceive them as requiring less skill, talent, and creativity, and assign them lower value \cite{horton_jr_bias_2023}. Correspondingly, many respondents remain doubtful that AI can truly create art \cite{xu2023everyone}.

Despite these concerns, AI is increasingly viewed as a tool for supporting human creativity. Acceptance of AI-assisted creation has grown in recent years \cite{hong2021artificial}. For example, a survey by Adobe found that three quarters of artists in the US, UK, Germany, and Japan would consider using AI as a creative assistant, though primarily for supporting tasks such as editing rather than core creative decisions \cite{anantrasirichai2022artificial}. Interviews similarly indicate that some artists see generative AI as expanding accessibility and enabling new forms of collaboration between humans and machines \cite{ali_studying_2023}.

\begin{figure*}[ht!]
\centering
  \includegraphics[width=.8\textwidth]{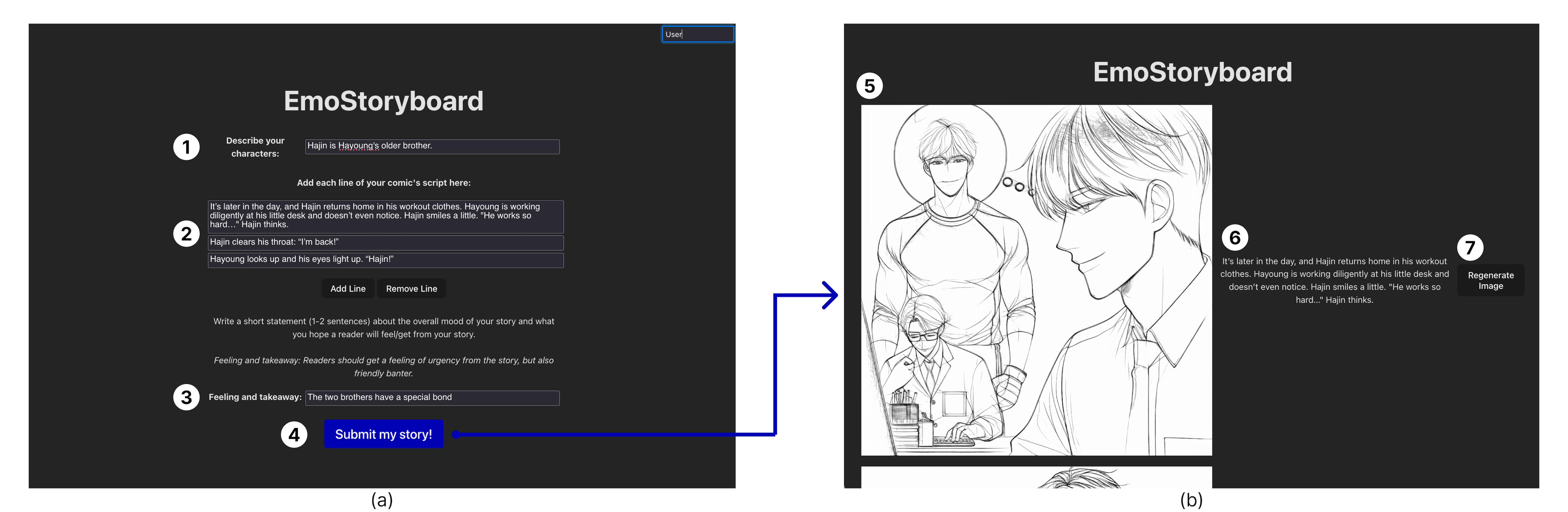}
  \caption{The user interface of EmoToon consists of two pages, (a) Comic Creation Page and (b) Storyboard Presentation Page. The
Comic Creation Page has (1) Character Description Input, (2) Comic Script Input, (3) Feeling and Takeaway Input, and (4) Submit Button. The Storyboard Presentation Page has (5) Generated Storyboard Panel, (6) Panel Script Text Box, (7) Regenerate Panel Button}
  \label{fig:interface}
  \vspace{-15pt}
\end{figure*}
\subsection{Human-AI Co-Creation Interfaces}

A growing body of work explores tools that support human creativity across domains such as music \cite{frid_music_2020}, data science \cite{zheng_telling_2022}, calligraphy \cite{haoran_magical_2023}, engineering \cite{yang_simulearn_2020}, animation \cite{subramonyam_taketoons_2018}, design ideation, and creative writing \cite{lee_coauthor_2022}. These systems extend creators' capabilities by combining AI strengths in efficiency and large-scale knowledge with human intuition, emotion, and personal experience \cite{wu_ai_2021}.

Some systems position AI as an active creative collaborator. For example, \textit{Drawing Apprentice} contributes unexpected sketches during collaborative drawing to spark new artistic directions \cite{davis_apprentice_2015}, while creative sketching systems encourage reinterpretation and iterative idea exploration \cite{karimi_relating_2019, karimi2020creative}. Other tools instead scaffold technical skills or reduce labor, such as \textit{ShadowDraw}, which displays reference images during sketching \cite{lee_shadowdraw_2011}, assistive painting tools that manage visual detail like color \cite{benedetti_painting_2014}, or automation systems like \textit{FlatMagic} that perform repetitive drawing tasks \cite{yan_flatmagic_2022}.

AI has also been explored as a co-creative partner in storytelling. Systems such as \textit{Talebrush} visualize emotional trajectories of narratives through interactive graphs \cite{chung_talebrush_2022}, while \textit{StoryDrawer} transforms children's stories into illustrative scenes \cite{zhang_storydrawer_2022}. More recent work explores end-to-end AI-assisted creative pipelines that lower the barrier to producing complex visual stories \cite{id8-huang}. However, while AI can support everyday creativity, transformative artistic expression often remains closely tied to human emotional and cultural understanding \cite{rafner_creativity_2023}. This highlights co-creativity as a promising paradigm that leverages complementary strengths of humans and AI \cite{wu_ai_2021,miller2019artist}.
Despite this growing ecosystem, relatively little work examines how co-creative tools can support \emph{emotionally expressive} visual storytelling, particularly for non-expert users. We therefore turn to literature on technological support for emotional visual expression.

\subsection{Technology for Supporting Emotional Visual Expression}

Emotion is central to both artistic creation and audience engagement \cite{de_dreu_hedonic_2008, gonzalez-zamar_emotional_2021}, yet conveying emotion visually remains challenging even for experienced artists \cite{ma_guidelines_2012}. Prior systems have explored computational approaches to mapping emotion to visual forms, though many focus on abstract or non-figurative representations \cite{wang2020design, krcadinac2015textual, xie2023creating}. While these approaches demonstrate how computational methods can encode affect, they do not address narrative techniques such as character gesture, camera framing, or scene composition that are central to emotional storytelling in comics \cite{cohn2016vocabulary, chavanne, stamenkovic}.

One notable exception is \textit{EmoG}, an AI-assisted sketching tool that generates emotional facial expressions for a drawn character \cite{shi_emog_2020}. However, its scope is limited to individual faces and the six-class basic emotion model, without incorporating narrative context or broader visual strategies such as body language, staging, or composition \cite{lu_shape_2012, mou_storyboard_2013, comic-spin,shi_emog_2020,al2023unique}.

\revision{

In contrast to systems that focus primarily on emotion recognition, \textsc{EmoToon} aligns with affective loop and emotion-as-interaction perspectives \cite{de_dreu_hedonic_2008}, treating user affect as a generative input rather than a recognition target.}
We address gaps with prior works through \textsc{EmoToon}, a technology probe that explores \textbf{AI-assisted comic storyboarding for conveying complex emotions in figurative form}. Our system provides suggestions that help \textbf{non-expert creators} translate emotional narrative intent into visual panels, and our user study evaluates its impact on (1) emotional expressiveness, (2) aesthetic quality, (3) perceived creativity and ownership, and (4) ease of learning storyboarding.

\section{EmoToon Technology Probe Design}

To empirically investigate users' creative expression during AI-assisted comic storyboarding, we first introduce the design of \textsc{EmoToon}, a fully-functioning technology probe that is available to users through a web application. \textsc{EmoToon} is a tool intended to guide the comic storyboarding process for non-experts. Our system takes in character descriptions, a comic script divided by panel, and an intended feeling and takeaway of the story (all as text inputs), and generates one storyboard panel per input panel. This probe was developed to help us better understand how AI assistance affects emotional expression in co-created comic storyboards.

\subsection{Visual Design Elements for Emotion Expression in Storyboarding}

\begin{quote}
    ``\textit{As the old proverb says,
`A picture is worth a thousand words.' If the storyboard
image can tell the story, then the writing of dialogues is
less important}'' \cite{mou_storyboard_2013}.
\end{quote}

It is well known that design elements can influence the way art is perceived \cite{riitijoki_expression_2021}. We begin the design of our technology probe by drawing on visual design theories known to convey emotions in the storyboarding process, focusing specifically on visual elements within the comic panel.
Through our review, we identify two key storyboard elements specifically tied to emotional expression: (1) composition \cite{chavanne,lu_shape_2012} and (2) facial and gestural character expression \cite{mou_storyboard_2013, comic-spin,shi_emog_2020,stamenkovic,al2023unique}.

Composition and character expressions have been studied extensively through empirical studies conducted on subjects' emotional responses to images \cite{zhang_analyzing_2011-1, freedberg_composition_2006, damiano2023anger}. 
Studies have shown that emotions are aroused by shapes present in images, finding that people felt higher valence emotions with rounder shapes and more simplistic image compositions \cite{lu_shape_2012} and compositional elements like harsh, sharp lines correlate directly to emotional effects like despair or sadness \cite{freedberg_composition_2006}. In our study, we define effective composition as the alignment \textbf{between the arrangement of visual elements within a storyboard panel and the emotions that are conveyed.}

When characters exhibit clear and relatable emotional expressions, it enhances readers' empathy and engagement with the story \cite{stamenkovic}. 
For example, animators will often push the gesture or facial expression of characters to efficiently convey an emotion. Our definition of effective character expression focuses on how \textbf{clearly facial expressions or gestures communicate the emotional states and intentions of characters.}



\subsection{EmoToon Design}
\paragraph{Emotion-driven input} 
We include a ``feeling and takeaway'' input (asking the author to explicitly describe the intended emotion and takeaway they hope one would get) to assist with the emotional element and anchors the entire generation pipeline in the artist's expressive intent. Storyboarding literature emphasizes that the emotional intent of a scene should guide the script and visual choices, meaning that the feeling to be conveyed is typically clarified before detailed storyboards are drawn \cite{glebas2012directing}, a workflow mirrored by our probe.

\paragraph{Idea Regeneration}
We added a panel regeneration function to support the exploratory and iterative nature of storyboarding, where artists frequently experiment and revise \cite{beiman2015prepare}. 



\paragraph{EmoToon System Pipeline}

\begin{figure}[t!]
\centering
  \includegraphics[width=\linewidth]{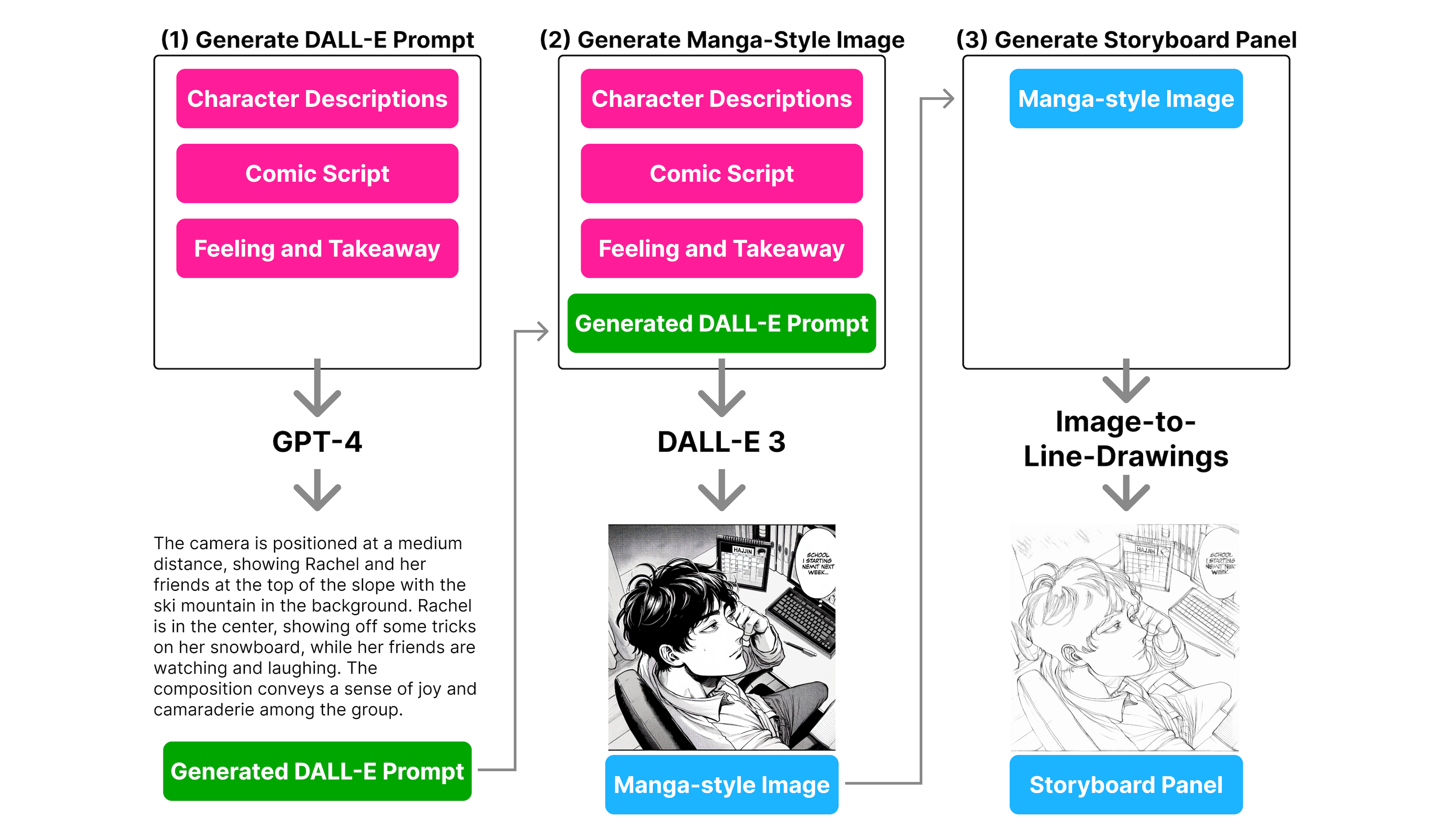}
  \caption{Storyboard generation pipeline, with example outputs.}
  \label{fig:imagegeneration}
  \vspace{-15pt}
\end{figure}

We construct the following steps in the \textsc{EmoToon} pipeline:
\begin{enumerate}
    \item Generating a DALL-E prompt with GPT-4: Based on the user's comic script, character descriptions, and intended emotion, we use a detailed prompt with GPT-4 to generate the textual DALL-E Prompt, including specific language about the \textbf{camera angle, panel composition, character expressions, and character poses}---in addition to the user-inputted emotion.
    \item Generating a Manga-style image with DALL-E: We combine the generated DALL-E Prompt with the user-inputted data (character descriptions, comic script, and feeling and takeaway) as part of a larger image generation prompt, which is then fed into DALL-E to obtain one manga-style image representing each panel of the user's comic script. 
    \item Converting Manga-style image to Storyboard Panel with the Image-to-Line-Drawings model: We run  the manga-style images outputted by GPT-4 through an Image-to-Line-Drawings model accessed via huggingface\footnote{\url{https://huggingface.co/spaces/awacke1/Image-to-Line-Drawings}}.
\end{enumerate}

\paragraph{User Interface}
The EmoToon user interface contains a Comic Creation Page (Fig.\ref{fig:interface} (a)) and a Storyboard Presentation Page (Fig.\ref{fig:interface} (b)). Users begin their interaction with EmoToon on the Comic Creation Page, where they input descriptions of the characters in their comic (Fig.\ref{fig:interface} (a,1)), the script of their comic divided by panel (Fig.\ref{fig:interface} (a,2)), and their intended feeling and takeaway (Fig.\ref{fig:interface} (a,3)), all as text. When users press the submit button (Fig.\ref{fig:interface} (a,4)), they are then taken to the Storyboard Presentation Page. On this page, users can view the generated storyboard panels (Fig.\ref{fig:interface} (b,5)), view or edit their panel scripts (Fig.\ref{fig:interface} (b,6)), and regenerate the storyboard image if desired (Fig.\ref{fig:interface} (b,7)).


\section{User Study}
To understand how people balance AI guidance with their own creativity in constructing personal storyboards, we conduct a user study with $N=25$ participants. Our main experimental manipulation is the use of \textsc{EmoToon} to guide storyboard drawing versus drawing completely \textsc{Freehand}. 

\subsection{Participants}
We recruited $N=25$ participants via mailing lists, focusing on adults with an interest in art. The sample covered a range of drawing and storytelling experience levels, but importantly \textbf{none were professional artists}. 
Basic demographics and self-reported art experience of our storytellers are summarized in Table \ref{tab:demographics}. 
\begin{table}[t]
    \centering
        \caption{Demographic and Art Experience of Storyboarding Participants who used \textsc{EmoToon}.}

    \footnotesize
    \resizebox{.8\linewidth}{!}{
    \begin{tabular}{|c|l|}
    \hline
    \textbf{Gender} & \begin{tabular}{l}
Female: $16 (64.00 \%)$ \\
Non-binary: $5 (20.00 \%)$ \\
Male: $4 (16.00 \%)$
\end{tabular} \\
    \hline
    \textbf{Age} & \begin{tabular}{l}
21.88  $\pm$ 3.10 \\
min: 18.00 \\
max: 29.00
\end{tabular} \\
    \hline
    \textbf{Ethnicity} & \begin{tabular}{l}
Asian: 20 (80.00\%) \\
Black or African American: 2 (8.00\%) \\
White: 1 (4.00\%) \\
Hispanic or Latino: 1 (4.00\%) \\
Mixed ethnicity: 1 (4.00\%)
\end{tabular} \\
    \hline
    \textbf{Drawing Skill Level} & \begin{tabular}{l}
Fair: 7 (28.00\%) \\
Good: $7 (28.00 \%)$ \\
Poor: $6 (24.00 \%)$ \\
Very good: $4 (16.00 \%)$ \\
Very poor: $1 (4.00 \%)$
\end{tabular} \\
    \hline
    \textbf{Storytelling Skill Level} & \begin{tabular}{l}
Fair: $10 (40.00 \%)$ \\
Good: $7 (28.00 \%)$ \\
Poor: 5 (20.00\%) \\
Very good: $3 (12.00 \%)$
\end{tabular} \\
    \hline
    \end{tabular}}
    \label{tab:demographics}
    \vspace{-10pt}
\end{table}
All participants were compensated \$20 for completing the study, which involved both creating storyboards and later evaluating others' storyboards.

\subsection{Study Design and Procedure}

\textbf{Pre-Intervention.} After providing consent and demographic information, participants joined a synchronous Zoom session. A researcher introduced the concept of storyboarding and demonstrated how to use the tool. Participants were then asked to write two short personal stories, each specifying a desired feeling and takeaway, and to create a 4-panel storyboard depicting an important moment in their life.

\textbf{Study Conditions and Tasks.} Participants completed two conditions: \textsc{EmoToon} and a \textsc{Freehand} baseline where they drew the storyboard from imagination. \revision{Critically, participants generated two completely separate narratives during the pre-intervention phase, assigning one distinct story to the \textsc{EmoToon} condition and the other distinct story to the \textsc{Freehand} condition. The pairing of stories and the execution order of conditions were fully counterbalanced and randomized across participants to guarantee that no single story prompt was repeated or illustrated across both workflows.} We used a \textsc{Freehand} condition to isolate the effect of the technology probe. More complex workflows (e.g., reference searching) were excluded because differences in prior experience and existing AI-assisted retrieval tools could introduce confounds. This study design prioritizes \emph{internal validity} over ecological realism, consistent with prior work on creativity support tools \cite{doshi_generative_2024, lee_shadowdraw_2011}.

\textbf{Post-Study.} After each task, participants completed a survey, followed by a 15-minute semi-structured interview after both conditions. In total, participants produced 50 storyboards (two per participant).

\subsection{Hypotheses} We designed our study to test the following hypotheses:

\begin{itemize}
  \item\textbf{[H1]} Users feel that \textsc{EmoToon} helps them better express the feeling of the story than drawing \textsc{Freehand}.
\item\textbf{[H2]} Users will be more satisfied with the aesthetic of their final drawings when using \textsc{EmoToon} vs drawing \textsc{Freehand}. 
\item\textbf{[H3]} Users will not perceive significant differences in creative freedom between using \textsc{EmoToon} and \textsc{Freehand} drawing. 
 \item\textbf{[H4]} \textsc{EmoToon} lowers the learning curve for storyboard drawing compared to \textsc{Freehand}. 
\end{itemize}

\begin{table*}[t]
    \centering
        \caption{Themes from Participant Interviews}

    \resizebox{.7\linewidth}{!}{
\begin{tabular}{l|l|l}
\hline
\textbf{Themes}                              & \textbf{Subthemes}                    & \textbf{Definitions}                                                          \\ \hline
{Design Assistance} & Ease of drawing              & Tool made drawing specific elements easier.                          \\
                                    & Help with layout             & Helped with overall page layout and flow.                            \\
                                    & Help with style              & Provided stylistic references that influenced the final product.     \\
                                    & Emotion                      & Tool helped with emotional output.                                   \\
                                    & Limited personal input       & Over-reliance on tool's output limited personal creativity.          \\ \hline
{Creative Influence} & Idea Generation       & Tool provided new composition and gesture ideas.                                       \\
                                    & Inspirational Background     & Suggested backgrounds that inspired new scenes.                      \\
                                    & Character Development        & Enhanced character portrayal through generated poses.                \\
                                    & Rigid Output                 & Felt constrained by tool's preset options.                           \\
                                    & Dramatic Scenes              & Imposed dramatic anime-style scenes that didn't fit user's vision.   \\ \hline
{Tool Satisfaction} & Creative Satisfaction & Satisfaction from generating new ideas and concepts.                                   \\
                                    & Ease of Use                  & Found the tool user-friendly and accessible.                         \\
                                    & Enhanced Productivity        & Felt more productive using the tool.                                 \\
                                    & Dissatisfaction with Outcome & Discontent with the final product not aligning with personal vision. \\ \hline
{Technical Challenges} & Output Error          & Errors and bugs during tool usage, such as generated text or image not matching input. \\
& Generation Time              & Tool took a long time to generate images.                            \\ \hline
\end{tabular}}
    \label{tab:interview-themes}
    \vspace{-10pt}
\end{table*}


\subsection{Data Collection and Analysis}\label{datacollection}
To answer \textbf{H1 - H4}, we specifically asked storytellers to self-rate (on a Likert scale of 1-7) the expressiveness of their storyboards (\textsc{Emotionally Expressive}, and \textsc{Art and Story Match}), aesthetic appeal (\textsc{Aesthetic}, \textsc{Character Expressions}, \textsc{Composition}), creativity (\textsc{Helpful for Exploration}, \textsc{Ease of Exploration}, \textsc{Final Storyboard Creative}, \textsc{Felt Creative}, \textsc{Degree of Freedom}, and \textsc{Ownership}), and learning curve of the tool  (\textsc{Learning}, \textsc{Ease of Use}, and \textsc{Confidence}).

\begin{figure}[t!]
    \centering
    \includegraphics[width=\linewidth]{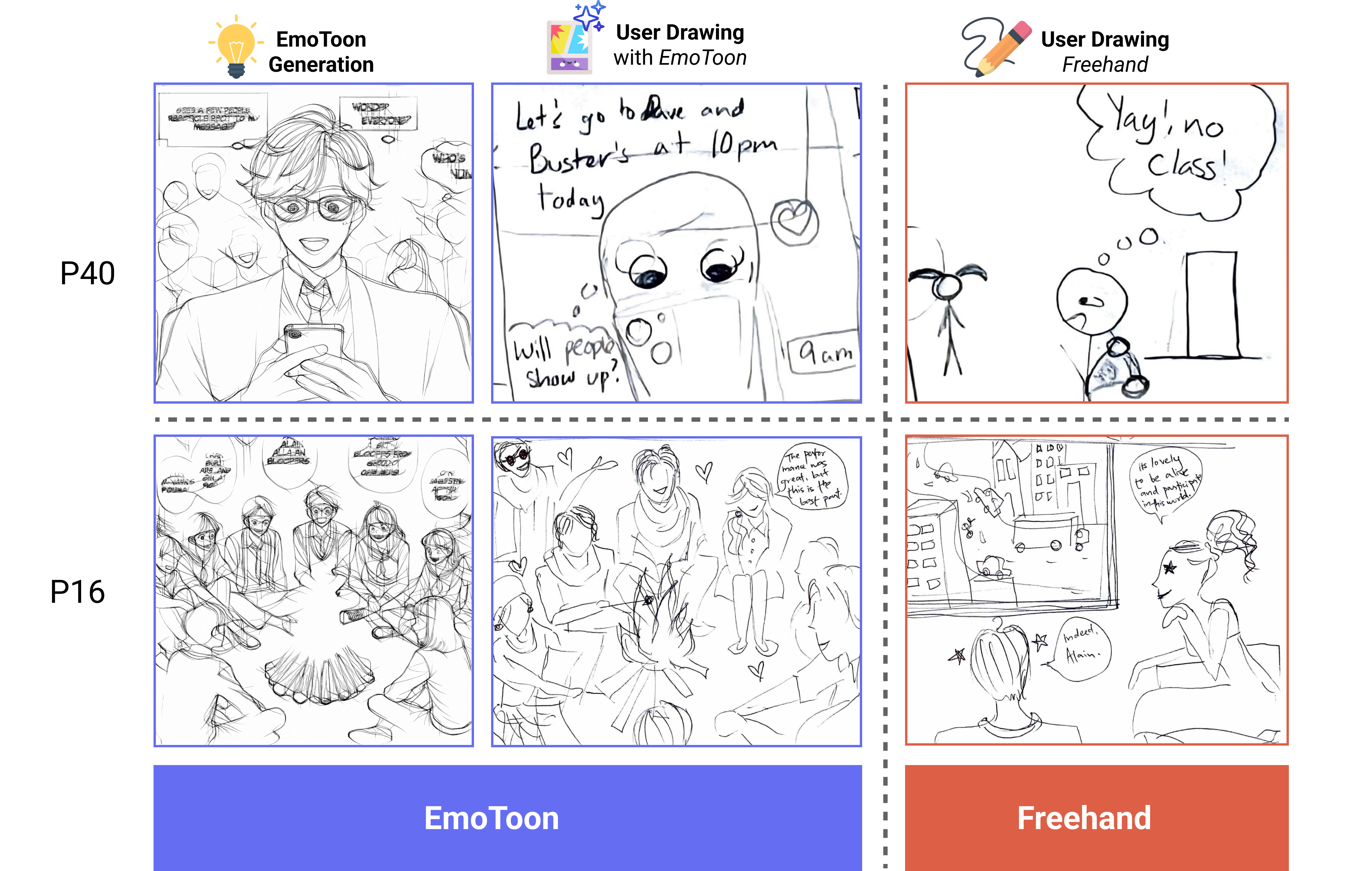}
    \caption{Example participant-drawn panels across \textsc{EmoToon} and \textsc{Freehand} conditions. }
    \label{fig:participant-examples}
    \vspace{-10pt}
\end{figure}

In addition to \textbf{survey, interaction, and artifact data, we also recorded and transcribed 15-minute interviews} with all participants after interacting with the tool. Our interview questions included \revision{participants' specific feedback regarding the interactive components and features of the EmoToon user interface (such as the panel regeneration mechanism),} as well as questions around how the tool aided their own emotional expression process or impacted their self-perceived creativity.
For interview and artifact analysis, two raters independently coded each interview utterance using the finalized themes and subthemes. We calculated Cohen's Kappa to measure inter-rater reliability and achieved substantial inter-rater agreement (where $\kappa = 0.67, 0.61 < \kappa < 0.80$ for interviews, $0.61 < \kappa < 0.80$ for artifacts) \cite{mchugh_interrater_2012}. Interview themes, subthemes, and their definitions are presented in the Results section (Table \ref{tab:interview-themes}), and example artifacts are shown in Fig.\ref{fig:participant-examples}.


\section{Results}

\begin{figure}[t!]
    \centering
    \includegraphics[width=\linewidth]{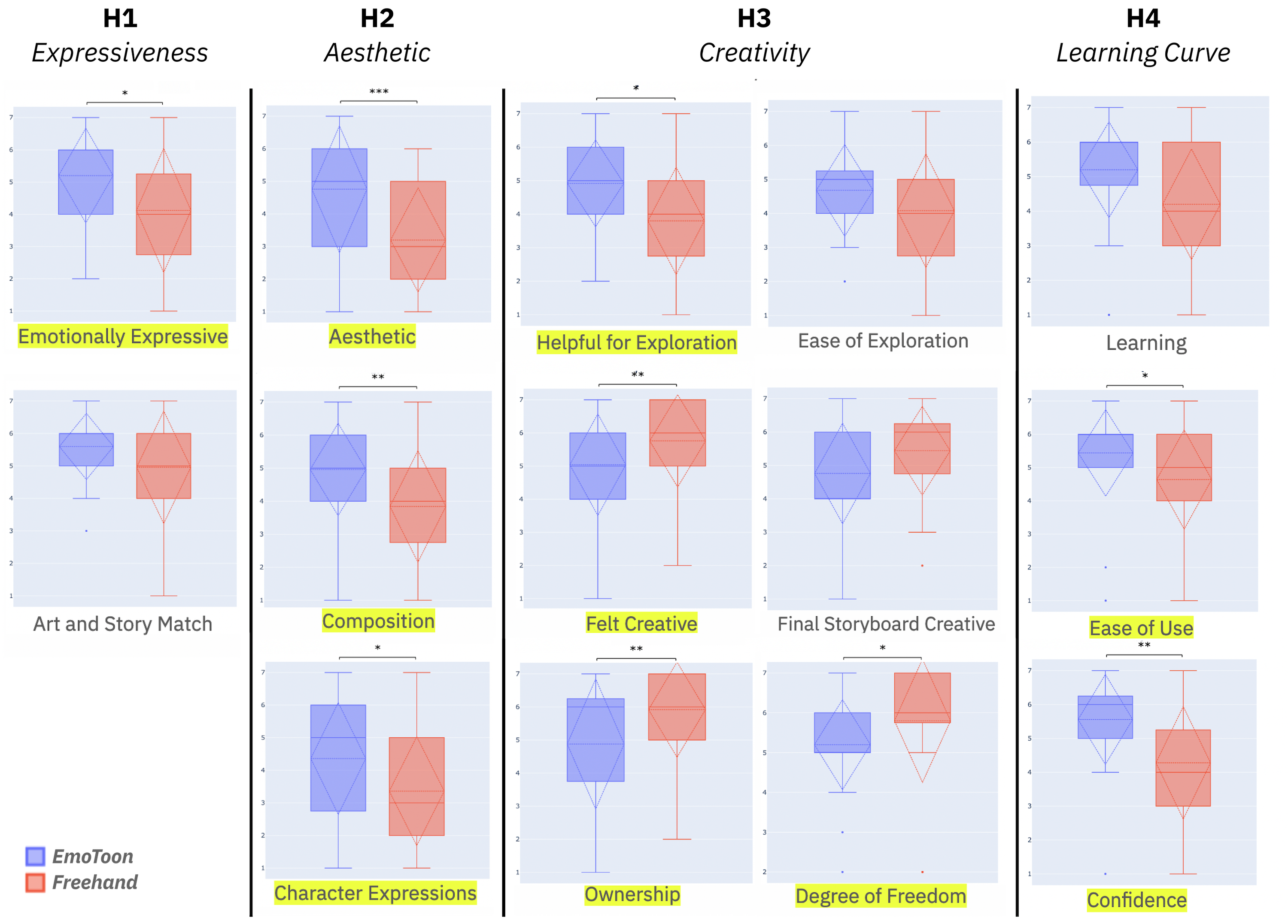}
    \caption{Quantitative metrics for effects on self-perceived emotional expression, aesthetic quality, creativity, and learning curve across \textsc{EmoToon} vs \textsc{Freehand} conditions. Significant differences are highlighted in yellow.}
    \label{allsurveys}
    \vspace{-10pt}
\end{figure}

\subsection{H1: Effects on Emotional Expression}


\subsubsection{Quantitative Insights}

We find that \textbf{\textsc{EmoToon} significantly improves self-perceived emotional expressiveness of the user's storyboard compared to \textsc{Freehand}}. First, users rated whether "the final storyboard is emotionally expressive" (\textsc{Emotionally Expressive}). EmoToon (M = 5.2, SD = 1.5) produces significantly higher scores ($W = 27, p = 0.01, N=25, r = 0.50$) than \textsc{Freehand} drawings (M = 4.12, SD = 1.96). However, we did not find statistically significant differences in whether writers believed their \textsc{EmoToon} storyboard better expressed the mood they intended than \textsc{Freehand} drawings (\textsc{Art and Story Match}).

\subsubsection{Qualitative Insights}

Interviews and artifact analyses indicate that \textsc{EmoToon} supported emotional expression primarily through \emph{design assistance} and \emph{creative influence}. Participants reported that generated references for facial expressions, gestures, and panel composition helped them visualize emotional states they might otherwise struggle to depict. As one participant explained, the system helped them express emotions they found difficult to imagine drawing themselves: \textit{``it was just hard for me to imagine how I would show my panic... so I think it helped me show that''} (P15).

Artifact analysis further suggests strong adoption of AI-generated emotional cues. In 23 of 25 storyboards, at least two panels incorporated emotional inspiration from generated examples, often through facial expressions, body language, or composition.

At the same time, participants noted limitations in emotional nuance and flow. Some felt the tool favored exaggerated or dramatic emotional styles that did not always match their intended tone, and several described inconsistencies in emotional progression across panels.

\subsection{H2: Effects on Aesthetic Quality}


\subsubsection{Quantitative Insights}

Writers who used \textsc{EmoToon} (M = 4.76, SD = 1.98) found their storyboards \textbf{significantly more \textsc{Aesthetic}} ($W = 15, p = 0.0007, N = 25, r = 0.68$) than drawings created \textsc{Freehand} (M = 3.2, SD = 1.63). 

\subsection{H3: Effects on Creativity}


\subsubsection{Quantitative Insights}

Users found \textsc{EmoToon} (M = 4.92, SD = 1.32) \textbf{significantly more helpful for exploring ideas} ($W = 32, p = 0.04, N = 25, r = 0.42$) than \textsc{Freehand} drawing (M = 3.8, SD = 1.63). While \textbf{\textsc{Ease of Exploration} was slightly higher for \textsc{EmoToon}} (M = 4.68, SD = 1.38) than \textsc{Freehand} (M = 4.08, SD = 1.71), these results were not significant.

However, \textbf{users did not find their \textsc{EmoToon} storyboards significantly more creative} (M = 4.76, SD = 1.53) than \textsc{Freehand} (M = 5.44, SD = 1.36), and \textbf{felt significantly less creative during \textsc{EmoToon} drawing} ($W = 18.5, p = 0.008, N = 25, r = 0.53$, M = 5.04, SD = 1.57) compared to \textsc{Freehand} drawing (M = 5.76, SD = 1.42).

Finally, \textbf{users felt significantly less \textsc{Ownership} and \textsc{Degree of Freedom}} (\textsc{Ownership}: $W = 14, p = 0.008, N = 25, r = 0.53$; \textsc{Degree of Freedom}: $W = 42, p = 0.05, N = 25, \revision{r = 0.39}$) when drawing with \textsc{EmoToon} (\textsc{Ownership}: M = 4.88, SD = 2.01; \textsc{Degree of Freedom}: M = 5.2, SD = 1.15) compared to \textsc{Freehand} (\textsc{Ownership}: M = 5.92, SD = 1.47; \textsc{Degree of Freedom}: M = 5.8, SD = 1.58).

\subsubsection{Regenerate Feature}

\begin{figure}[h]
    \centering
    \includegraphics[width=.5\linewidth]{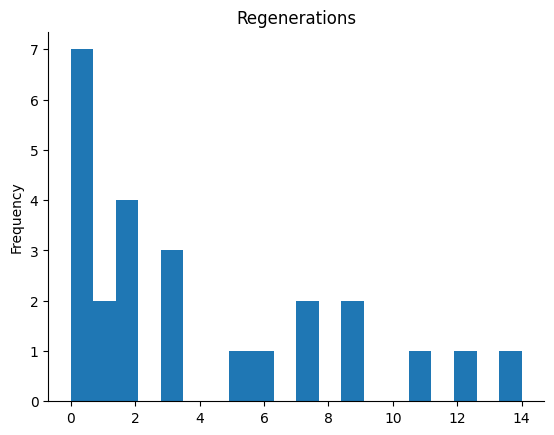}
    \caption{Distribution of frequency for use of the \textsc{EmoToon} regeneration tool.}
    \label{fig:regen}
\end{figure}

For creative iteration, participants varied widely in their use of the regenerate feature (Figure~\ref{fig:regen}). Some repeatedly regenerated outputs to explore alternatives, while others rarely used the feature. For example, one participant described the tool as enabling rapid iteration: \textit{``you can keep iterating until you find something that either gives you a really good idea, or that you like more."}  (P22). These behaviors suggest that regeneration supported exploration for some users while others preferred working with a single generated reference.

\subsubsection{Qualitative Insights}

Qualitative findings indicate that \textsc{EmoToon} expanded users' idea space while also introducing tensions around authorship. Participants frequently described generated examples as useful starting points for brainstorming and visual exploration. 
However, others reported feeling constrained by initial AI outputs, which sometimes anchored their thinking or reduced their sense of ownership. For example, P25 mentioned that they felt \textit{``stuck with it, my mind was kind of stuck with only seeing that one thing. And so I couldn't really be that creative with the bigger picture details.''} Despite this tension, several participants actively modified or rejected generated suggestions, indicating that creativity often manifested through adaptation rather than direct adoption. 

\subsection{H4: Effects on Learning Curve}

\subsubsection{Quantitative Insights}

We report perceived learning curve across \textsc{Learning}, \textsc{Ease of Use}, and \textsc{Confidence}. Users reported higher \textsc{Learning} scores for \textsc{EmoToon} (M = 5.2, SD = 1.41) than \textsc{Freehand} (M = 4.2, SD = 1.63), though this difference was not significant.

Users found \textsc{EmoToon} (M = 5.44, SD = 1.33) \textbf{significantly easier to use} ($W = 33.5, p = 0.02, N = 25, \revision{r = 0.46}$) than \textsc{Freehand} drawing (M = 4.64, SD = 1.52). Writers also felt \textbf{significantly more confident} using \textsc{EmoToon} ($W = 18.0, p = 0.005, N = 25, r = 0.56$, M = 5.56, SD = 1.36) than drawing \textsc{Freehand} (M = 4.28, SD = 1.70).

\subsubsection{Qualitative Insights}

Interview data aligns with these findings: 19 of 25 participants reported that \textsc{EmoToon} simplified aspects of storyboarding. Participants described the system as reducing technical barriers such as drawing poses, perspectives, and consistent environments across panels. One participant explained that the tool made technically challenging drawings easier: \textit{``I struggle a lot with poses and angles... so it made the technical parts a lot easier''} (P18). Participants also reported increased confidence through the availability of visual references, which allowed them to experiment with styles and compositions they might not otherwise attempt.
\section{Broader Discussion, Limitations, and Future Directions}

Our study examined how an emotion-driven, AI-assisted storyboarding tool can support non-expert creators in visualizing personal narratives. Overall, \textsc{EmoToon} improved perceived emotional expressiveness (\textbf{H1} partially accepted) and aesthetic quality (\textbf{H2} accepted) relative to \textsc{Freehand} drawing, and increased ease of use and confidence (\textbf{H4} partially accepted), but came with costs to perceived creativity and ownership (\textbf{H3} rejected). These results suggest a core trade-off: emotion- and composition-rich references can help novices externalize affect and produce more polished storyboards, yet the same high-level structural proposals may anchor authorship and reduce perceived agency.


A key contribution of \textsc{EmoToon} is treating emotion as a driver of the co-creative workflow rather than only an evaluation target. Conditioning generation on a ``feeling and takeaway'' helped participants translate emotional intent into concrete visual choices spanning facial expression, gesture, and composition, extending prior work that emphasizes character-centric cues or single-image emotion transfer \cite{shi_emog_2020, lu_shape_2012}. In storyboarding, however, emotion is constructed across framing and panel progression, not only within a face \cite{nguyen_comic_2017}. Our findings motivate systems that treat emotional goals as \emph{dynamic}: interfaces could support lightweight per-panel affect tags and rapid adjustment when intentions shift, and use interaction traces (e.g., repeated regeneration or rejection) as signals of misalignment that prompt clarification in an iterative workflow \cite{sawyer_iterative_2021}.

Comics and storyboards impose sequential constraints that current co-creative paradigms often under-address. Participants reported that AI could assist per-panel emotion but sometimes disrupted \emph{emotional coherence} across panels via abrupt tone shifts or melodramatic ``spikes.'' This suggests future storyboard tools should reason about emotion \emph{sequentially}, coupling text, layout, and affect across panels, potentially enabled by data-centric approaches using large-scale comic datasets with artist consent \cite{nguyen_comic_2017, yang_emogen_2024}. Interaction designs could visualize an emotional arc and offer compositional alternatives (e.g., inserting a wider interstitial shot) to smooth or intensify progression, extending sentiment-arc ideas from text to visual narrative \cite{yeh_story_2025}. More broadly, composition and shape design remain undersupported despite being central levers of affect; tools could provide ``composition palettes'' that let users explore framing, panel rhythm, and spatial weight as an \emph{abstract} form of creativity support \cite{reis_using_2008, ma_guidelines_2012}.

Consistent with prior co-creation work, participants experienced tension between exploratory benefits and reduced ownership \cite{ali_studying_2023, horton_jr_bias_2023, davis_apprentice_2015, chung_promptpaint_2023, hwang_it_2025}. Notably, even though \textsc{EmoToon} outputs were sketch-like references rather than polished renders, users still reported lower ownership and freedom, suggesting that \textbf{high-level structural decisions (composition, pose) may shape authorship perceptions as much as surface-level polish}. Discrete proposal-based interaction may further amplify anchoring relative to more continuous steering interfaces \cite{chung_promptpaint_2023}, raising risks of fixation and reduced divergence \cite{chrysikou_functional_2016, tholander_design_2023, benedetti_painting_2014}. Future work should compare AI-generated references to conventional references (e.g., photos or other artists' work) and develop measures that better capture ownership in visual, spatial creation beyond attribution by ``idea proportion'' \cite{hwang_it_2025}.

Finally, \textsc{EmoToon} lowered barriers for novice creators by increasing ease of use and confidence, aligning with ``low floor, high ceiling'' creativity support \cite{shneiderman_creativity_2007}. However, participants also questioned whether AI support translated into actual learning versus outsourcing. Concretely, future systems could offer ``guided'' modes with explanatory overlays and prompts to modify references, and ``minimal'' modes that restrict AI to thumbnails, composition alternatives, or mood boards for experienced artists \cite{lima_laypeoples_2024, zhang_perceptions_2024, mikalonyte_can_2022, ali_studying_2023}. Transparency cues that mark AI-proposed elements may further support reflection and attribution norms in co-creative workflows \cite{ali_studying_2023, hwang_it_2025}.





\section{Conclusion}
In this work, we presented \textsc{EmoToon}, a technology probe based on visual design elements that influence emotion, in order to empower non-expert users to create their own comic storyboards. Our user study with 25 users of \textsc{EmoToon} showed that our system aided in self-perceived emotional expression, aesthetic quality, creativity exploration, and increased user confidence compared to drawing \textsc{Freehand}. However, we also found that participants felt significantly lower ownership and freedom when using AI assistance in storyboard creation, and suggest future directions in varying abstraction of artistic support provided by systems to increase the user's own sense of creative voice for this medium.
Our work paves future directions in understanding the strengths and limitations of AI tools for empowering users with diverse art and storytelling backgrounds to create emotionally impactful storyboards.

\section*{Ethical Impact Statement}

Because our system engages with emotional storytelling and generative media, several ethical considerations arise regarding human subjects research, privacy, potential misuse of generative technologies, and the limits of generalizing our findings.

Firstly, the study involved participants creating storyboards based on personal experiences and emotional narratives. The research was conducted under institutional ethics oversight (anonymized for review), and all participants provided informed consent prior to participation. Participants were informed that they could skip questions, change their story topic, or withdraw from the study at any time without penalty. Because recalling personal experiences can sometimes involve sensitive emotions, we emphasized that participants should only share stories they felt comfortable discussing and avoided prompts that required disclosure of highly sensitive or identifying information.

The study generated several forms of potentially sensitive data, including participant drawings, interaction logs, generated storyboard references, and interview transcripts. These materials may indirectly reveal personal experiences or identities. To mitigate privacy risks, we limited data collection to information necessary for analysis, removed identifying details during transcription and reporting, and stored all data securely with access restricted to the research team. Any artifacts reproduced in publications are screened to avoid revealing identifying information.

As the system relies on generative image models, there are also broader societal concerns related to generative media. While \textsc{EmoToon} is designed as a creativity support tool for ideation rather than photorealistic generation, similar technologies can be misused for deceptive media or non-consensual depictions. Additionally, generative models may reflect biases present in their training data, potentially reinforcing stereotypical portrayals of emotions, identities, or social roles. Our work does not attempt to resolve these broader challenges but highlights them as important considerations for future systems that build on generative storytelling technologies. 

\revision{Finally, we recognize that our participant sample and study design limits the generalizability of our findings. Norms of emotional expression, visual storytelling, and comic conventions vary significantly across cultures, which could influence how diverse populations interact with an emotion-driven generation tool. Additionally, choice of a \textsc{Freehand} drawing baseline prioritizes internal validity by avoiding confounds related to individual differences in reference search habits. Consequently, our claims regarding \textsc{EmoToon}'s benefits are framed relative to unassisted drawing rather than typical creative practice involving reference gathering, which may under-represent existing artists' diverse workflows. Future work should incorporate external assessments of emotional and compositional elements to thoroughly validate these self-reports, as well as ensuring a diverse, cross-cultural participant base in order to validate broader applicability.}

\bibliographystyle{IEEEtran}
\bibliography{sample-base}

\end{document}